%% file: paper_la.tex
\documentclass[twocolumn,showpacs,aps,prl,superscriptaddress]{revtex4}

\usepackage{graphicx}
\usepackage{dcolumn}
\usepackage{amsmath}
\usepackage{epsfig}

\input pubboard/babarsym
\input symbols

\newcommand{\BABARPubYear}    {01}
\newcommand{\BABARPubNumber}  {22}
\newcommand{\SLACPubNumber} {9096}

\def\figurebox#1#2#3{%
    \def\arg{#3}%
    \ifx\arg\empty
    {\hfill\vbox{\hsize#2\hrule\hbox to #2{\vrule\hfill\vbox to #1{\hsize#2\vfill}\vrule}\hrule}\hfill}%
    \else
    {\hfill\epsfbox{#3}\hfill}%
    \fi}

\def\Dstarp   {\ensuremath{D^{*+}}\xspace}
\def\Dstarm   {\ensuremath{D^{*-}}\xspace}

\def\dm {\Delta m_d}

\def\delz {\Delta z}
\def\delt {\Delta t}

\def\dz {\ifmath{\Delta z}}

\def\re_eb {\ifmath{Re(\varepsilon_B)}}
\def\gevsq  {\ifmath{\mbox{\,Ge\kern -0.08em V}^2}} 

\begin{document}

\preprint{\babar-PUB-\BABARPubYear/\BABARPubNumber}

\begin{flushleft}
\babar-PUB-\BABARPubYear/\BABARPubNumber\\
SLAC-PUB-\SLACPubNumber\\[10mm]
\end{flushleft} 

\title{
{\large \bf Measurement of the {\boldmath{\Bz-\Bzb}} Oscillation Frequency with 
Inclusive Dilepton Events}
}

\input pubboard/authors_0122

\date{\today}
\begin{abstract}
The \Bz-\Bzb\ oscillation frequency has been measured with a sample of 
23 million \BB\ pairs collected with the \babar\ detector at the 
\pep2\ \abf\ at SLAC. In this sample,
we select events in which both $B$ mesons decay 
semileptonically and use the charge of the leptons to identify the flavor of 
each $B$ meson. A simultaneous fit to the decay time difference distributions 
for opposite- and same-sign dilepton events gives 
$ \dm = 0.493 \pm 0.012\mbox{ (stat)}\pm 0.009\mbox{ (syst)} $ ps$^{-1}$. 
\end{abstract}


\maketitle

%
%

A precise measurement of the \Bz-\Bzb\ oscillation 
frequency $\dm$ is of fundamental importance as a direct measure of
the Cabibbo-Kobayashi-Maskawa (CKM) matrix element $|V_{td}|$ \cite{bib:CKM}. 
When combined with measurements 
of the \Bs-\Bsb\ oscillation frequency, it provides a stringent constraint 
on the Unitarity Triangle of the CKM matrix. 

In this Letter, we present a measurement of the time dependence of \Bz-\Bzb\ mixing 
using data collected with the \babar\ detector at the \pep2\ asymmetric energy 
\epem\ collider operated at or near the \FourS\ resonance. 
The data sample, recorded in the years 1999-2000,  
corresponds to an integrated luminosity of 20.7\invfb\ on the \FourS\ resonance 
({\it{on-resonance sample}}), and 2.6\invfb\ collected 40\mev\ below the \FourS\ 
resonance ({\it{off-resonance sample}}). 
\BB\ pairs from the \FourS\ decay move along the high-energy beam direction ($z$) 
with a nominal Lorentz boost 
$\langle \beta \gamma \rangle = 0.55$.
Therefore, the two \B\ decay vertices are separated by 
about $260 \mum$ on average. 

The measurement technique is based on the identification of 
events containing two leptons from semileptonic decays of $B$ mesons. 
The flavor of the $B$ mesons at the time of their decay is
determined or ``tagged'' by the charge of the leptons. 
Thus, for \FourS\ resonance decays into \BzBzb\ pairs, 
neglecting backgrounds, opposite-sign ($+$) and same-sign ($-$) lepton 
pairs correspond to unmixed and mixed events, respectively.
Because the \BzBzb\ pair is in a coherent $P$-wave state,
the time evolution of the \B\ mesons is a function of the proper time difference
$\delt$ between the two \B\ decays:
\begin{equation*}
{\cal{S}}_{\pm}(\delt; \dm) = \frac{e^{-|\delt|/\tau}}{4\tau} 
(1 \pm \cos{\dm \delt}),
\end{equation*}
where $\tau$ is the \Bz\ lifetime ($\Delta\Gamma=0$ is assumed). 
The corresponding time-dependent asymmetry is 
$ ({\cal{S}}_+(\delt) - {\cal{S}}_-(\delt))/
({\cal{S}}_+(\delt) + {\cal{S}}_-(\delt))= \cos{\dm \delt}$. 

This simple picture is modified by the effects of detector resolution 
and the presence of backgrounds. 
The most important background, about 50\% of the sample, is due to 
\BpBm\ events, which are not removed by the event selection critieria. 
The fraction of \BpBm\ events is 
determined from the data itself in order to reduce systematic uncertainties. 
Other non-negligible backgrounds are leptons 
from the $b\rightarrow c\rightarrow \ell$ decay chain ({\it{cascade decays}}), which are 
also the main source of wrong tags, and hadrons that are misidentified as leptons. 
Signal and background probability density functions (PDF) for opposite- and same-sign 
events are included as additional terms in the full PDF. The corresponding likelihood function, 
combining opposite- and same-sign dilepton events, is maximized to determine $\dm$. 

%
%
The \babar\ detector is described in detail elsewhere~\cite{bib:babar}. 
Charged particle tracking is provided by a 5-layer, double-sided silicon vertex tracker 
(SVT) and a 40-layer drift chamber (DCH), both operating inside a 1.5-T super-conducting 
solenoidal magnet. 
The CsI(Tl) electromagnetic calorimeter (EMC) 
detects photons and electrons.  
Particle identification is provided by a ring-imaging Cherenkov
detector (DIRC) and specific ionization measurements \dedx\ in the DCH.  
Muons are identified with the instrumented flux return (IFR), segmented to contain 
resistive plate chambers. 

%
%

Events are selected by requiring more than 5 reconstructed 
charged tracks, at least 3 of which must originate from the interaction region
and be reconstructed in the DCH. 
Continuum 
the normalized second Fox-Wolfram moment~\cite{bib:FoxW} be less than 0.4 and 
the event aplanarity be greater than 0.01. Two-photon events, as well as 
residual radiative Bhabha events with a large amount of missing energy,
are rejected by requiring the invariant mass squared of the event 
to be greater than 20\,GeV$^2/c^4$. 

Electrons are selected by requirements 
on the ratio of the energy deposited in the EMC to the momentum measured in the 
DCH, the lateral shape of the energy deposition in the EMC, and 
\dedx\ in the DCH. 
Muons are identified on the basis of the energy in the EMC, as well as 
the strip multiplicity, track continuity and penetration depth in the IFR. 
Lepton candidates consistent with the kaon hypothesis as measured in the DIRC are 
rejected. 
Electron (muon) selection efficiencies and misidentification
rates at high momentum
are about 92\% (75\%) and 0.15\% (3\%), respectively.

Photon conversions are rejected by 
combining each electron candidate with all other oppositely-charged
electron candidates 
in the event, selected with looser criteria, and applying requirements on the 
invariant mass and distance of closest approach of the pair
in the transverse plane and along the beam direction. 
Leptons from \jpsi\ and $\psi(2S)$ decays are identified by pairing 
them with all other oppositely-charged candidates of the same lepton species, 
selected with looser criteria, and rejecting the whole event if any combination 
has an invariant mass within the \jpsi\ or $\psi(2S)$ mass regions. 

Events with at least two leptons are retained and the two highest momentum leptons in 
the \FourS\ rest frame are used in the following. 
%
%

The two lepton tracks and a beam spot constraint are used in a vertex fit 
to find the primary vertex of the event in the transverse plane. 
The positions of closest approach of the two tracks to this vertex in the transverse plane 
are computed and their $z$ coordinates are denoted as $z_1$ and $z_2$, where 
the subscripts 1 and 2 refer to the highest and second highest momentum leptons in the 
\FourS\ rest frame. The difference $\delz$ is defined as $\delz = z_1 - z_2$. 
The time difference $\delt$ is computed from $\delz$ and the nominal boost as 
$\delt = \delz / \langle \beta \gamma \rangle c$. 

The modeling of the resolution function ${\cal{R}}$ is a crucial element 
of the $\dm$ measurement. To improve the $\delz$ (and therefore $\delt$) resolution, 
reduce the fraction of incorrectly measured tracks, and minimize related 
systematic uncertainties, charged tracks are required to satisfy the 
following criteria. Lepton candidates must have a distance of closest 
approach (DOCA) with respect to the 
nominal beam position of less than 1\cm\ in the transverse plane and 
less than  6\cm\ along the beam direction, at least 12 hits in the DCH, 
at least four $z$-coordinate hits in the SVT, a momentum in the range 
between 0.7 and 2.5\gevc in the \FourS\ rest frame and between 
0.5 and 5.0\gevc in the laboratory frame, and a polar angle in the range 
between 0.5 and 2.6 radians in the laboratory frame. 
The total error on $\delz$, computed on an event-by-event basis, is required 
to be less than 175\mum. 
The vertex fit constrains the lepton tracks to originate from 
the same point in the transverse plane, thereby neglecting the non-zero flight
length for \B\ mesons. 
As a consequence, the $\delz$ resolution function is 
$\delz$ dependent, becoming worse at higher $\vert\delz\vert$. 
Neglecting this $\delz$ dependence 
introduces a small bias, discussed below. 
Vertex studies with leptonic $J/\psi$ decays show 
that the $\delt$ resolution function for signal 
events can be appropriately modeled with the sum of three Gaussian distributions
in both data and Monte Carlo simulation. 

%
%
The separation between direct leptons and background from cascade decays 
is achieved with a neural network that combines five discriminating variables 
for each event and provides two outputs, one for each lepton, 
chosen to vary between 0 for cascade leptons and 1 for direct leptons. 
The discriminating variables are the momenta of the two leptons, 
their opening angle, the total visible energy, and the 
missing momentum of the event. All variables are computed in the \FourS\ rest frame. 
The first two variables are very powerful in discriminating between direct and
cascade leptons. The third efficiently removes 
direct-cascade lepton pairs from the same $B$ decay
and further reduces contributions from photon conversions. 
An optimization study based on the minimization of the total error on $\dm$ leads to 
the requirement that both neural network outputs be greater than 0.8.

%
%
The numbers of selected on-resonance and off-resonance events are 99010 and 
428, respectively. 
The combined effect of all requirements gives a direct 
dilepton purity and efficiency of about 83\% and 9\%, respectively, 
based on Monte Carlo simulation. 
Semileptonic $B$ decays in the Monte Carlo simulation have been modeled separately 
for each charm meson involved. 
A parametrization of HQET form factors \cite{bib:CLEOform} is used for 
$B \rightarrow D^* \ell \nu$, the Goity-Roberts model \cite{bib:GoityRoberts} is used 
for $B \rightarrow D^{(*)}\pi\ell\nu$, while the ISGW2 model 
\cite{bib:ISGW2} is used for $B \rightarrow D \ell \nu$ and 
$B \rightarrow D^{**} \ell \nu$. 
The measured branching fractions for decays to $D^{**}$ and $D^{(*)}\pi$ 
states are fixed to their world averages~\cite{bib:PDG2000}
and unmeasured processes have rates
that are inferred on the basis of isospin arguments.  
Events from \BB\ decays are grouped in three topologies, each of which 
is assigned its own PDF with different $\delt$ dependence and tagging properties. 

%
%
{\it{Direct dilepton}} events are described by the convolution of an oscillatory term 
for neutral $B$ decays, or an exponential function for charged $B$ decays, 
with the resolution function ${\cal R}$:
\begin{equation*}
{\cal{S}}^{n(c)} = {\frac{e^{- |\delt|/\tau_{n(c)}}}{4\tau_{n(c)}}}
(1 \pm D^{n(c)}_{\rm sig} \xi_{n(c)})\otimes {\cal{R}}
\end{equation*}
for neutral (${n}$) and charged (${c}$) events, 
where $\tau_{n(c)}$ is the $B$ meson lifetime, 
$\xi_{n}=\cos\dm\delt$ and $\xi_{c}=1$, and 
$D^{n (c)}_{\rm sig} \approx 0.95$ are correction factors 
that account for the (small) fraction of wrongly tagged direct 
dilepton events. 
These events are due to hadrons 
from the $B$ vertex that are misidentified as leptons 
or leptons from the decay of resonances ({\it{e.g.}}, events where 
only one lepton comes from a $J/\psi$) 
produced at the $B$ vertex. 
Both of these sources give almost random tagging and,
in the absence of such events, $D^{n (c)}_{\rm sig}$ would be exactly 1. 
A small fraction of events where a lepton originates from the 
$b\rightarrow \tau^-\rightarrow \ell^-$ decay chain have the same charge 
correlation as signal events and are also included 
in the signal topology. Neglecting 
the $\tau$ lepton lifetime introduces a negligible 
bias on the $\dm$ measurement. 

%
%
{\it{Opposite $B$ cascade}} (OBC) events, 9\% of the selected sample, contain 
one lepton from a $b\rightarrow \ell$ decay and the other from the 
$b\rightarrow c\rightarrow \ell$ decay chain of the companion $B$ meson. 
These events are the main source of wrong tags. 
Their PDFs are modeled by the convolution of $\delt$-dependent terms 
of a form similar to the signal
with a resolution model that takes into account the effect of the  
charmed meson lifetimes by convoluting 
three Gaussians with 
a single-sided exponential decay distribution.
Since both short-lived $D^0$ and $D_s$, 
and long-lived $D^+$ mesons are involved in 
cascade decays, the global OBC PDFs are written as  
\begin{equation*}
{\cal{C}}_{\rm OBC}^{n(c)} = {\frac{e^{- |\delt|/\tau_{n(c)}}}{4\tau_{n(c)}}}  
\sum_{i} f^{n(c)}_{i} (1 \pm D^{i,n(c)}_{\rm OBC}\xi_{n(c)}) \otimes {\cal{R}}^{i}_{\rm OBC}
\end{equation*}
where the index $i$ runs over the short- 
and long-lived charm meson components. 
This parameterization of OBC events significantly reduces the related systematic uncertainty.  
The two resolution functions ${\cal{R}}^i_{\rm OBC}$ allow for different
effective charm lifetimes and parameters of the three Gaussians, 
since the resolution function depends on the $B$ and $D$ flight lengths.
Due to the different decay processes involved, the relative fractions $f_i^{n (c)}$ 
of short- and long-lived charm mesons are also different in neutral and charged $B$ 
events. 
If particle identification were perfect and cascade 
leptons originated only from the $b\rightarrow c\rightarrow \ell^+$ 
process, then flavor tagging would always be wrong and the factors 
$D^{i, n (c)}_{\rm OBC}$ would be exactly $-$1. 
Hadron misidentification (PID) and resonance decays, as well as leptons originating from the 
$b \rightarrow c\overline c(\rightarrow \ell^-)s$ chain, give a fraction of 
right tags (15\%) even in the OBC topology. These two processes have been 
factorized by writing 
$D^{i, n (c)}_{\rm OBC}=D^{i, n (c)}_{\rm PID}\cdot
D^{i, n (c)}_{b \rightarrow c\overline c s}$ and assuming 
no correlation between the two terms. 

%
%
{\it{Same $B$ cascade}} (SBC) events, 4\% of the selected sample, 
contain two leptons from a single \B\ meson, obtained
via the decay chain $\b \rightarrow c \ell^- \overline \nu$, 
with $c \rightarrow x \ell^+ \nu$. 
SBC events are insensitive to mixing and,  
in the case of perfect particle identification and 
in the absence of resonances, would always give opposite-sign 
leptons. The PDFs are 
\begin{equation*}
{\cal{C}}_{\rm SBC}^{n (c)}={\frac{e^{- |\delt|/\tau^{n (c)}_{\rm SBC}} 
}{4\tau_{\rm SBC}^{n (c)}}} (1 \pm D^{n (c)}_{\rm SBC}),
\otimes \cal R
\end{equation*}
where $\tau^{n (c)}_{SBC}$ are effective lifetimes and $D^{n (c)}_{\rm SBC}$ 
are corrections for wrong tags in the SBC topology. The resolution $\cal R$ 
is taken to be the same as for signal events, 
with no significant bias on the final result. 

%
%
A small residual background remains (0.3\% of the 
total sample) where both leptons are from an unrecognized 
$J/\psi$ decay. 
These are described by a term $\Psi=\delta(\delt)\otimes{\cal{R}}$, whose 
normalization is obtained from simulation. 
Events where one lepton originates from a cascade decay and the other 
from a $B$ decay to $\tau$ or to a resonance, 
and events where both leptons come 
from cascade decays, (0.3\% of the total sample) are assigned the OBC event topology 
with no significant bias on $\dm$. 

The fraction $f_{\rm cont}= 3.4\%$ and $\delt$ dependence 
of the continuum background are determined from off-resonance data. The
$\delt$ dependence is parameterized for opposite- and same-sign 
leptons as ${\cal{Q}}_{\pm}=\tau_{\rm cont}^{-1}e^{- \tau_{\rm cont} |\delt|} f_{\pm}$, 
with $f_+ + f_- = 1$.

%
%
The full likelihood function is the product of likelihoods for opposite- 
and same-sign events, which can be schematically written as
{\allowdisplaybreaks
\begin{eqnarray*}
{\cal{L}} &=& (1-f_{\rm cont})(1-f_{J/\psi})[\\ \nonumber 
&           & (1-f_{c})( f_{\rm sig}^{n}{\cal{S}}^{n} 
        + f_{\rm OBC}^{n} {\cal{C}}_{\rm OBC}^{n} + f_{\rm SBC}^{n} {\cal{C}}_{\rm SBC}^{n}) 
        + \\ \nonumber
&           & + f_{c}(f_{\rm sig}^{c} {\cal{S}}^{c} 
        + f_{\rm OBC}^{c} {\cal{C}}_{\rm OBC}^{c} + f_{\rm SBC}^{c} {\cal{C}}_{\rm SBC}^{c})]
        + \\ \nonumber
&           & + (1-f_{\rm cont})f_{J/\psi} {\Psi} + f_{\rm cont} {\cal{Q}},\nonumber 
\end{eqnarray*}
}where the $J/\psi$ term and its relative abundance $f_{J/\psi}$ 
are present for opposite-sign events only, and
$f_{\rm sig}^{n (c)}=(1-f_{\rm OBC}^{n (c)}-f_{\rm SBC}^{n (c)})$. 
The fraction $f_{c}$ of charged $B$ events in the selected sample and 
the OBC fraction $f_{\rm OBC}^{n}$ in neutral $B$ events
are extracted from the fit. The OBC fraction $f_{\rm OBC}^{c}$ in charged $B$ events 
is scaled with $f_{\rm OBC}^{n}$ according to the value of the ratio 
$f_{\rm OBC}^{c}$/$f_{\rm OBC}^{n}$ determined with the Monte Carlo simulation. 
The SBC fractions are computed for simulated events and fixed in the fit.
The various parameters for the OBC resolution functions
are taken from a fit to Monte Carlo events. 
The factor 
$D^{c}_{\rm PID}(D^0,D_s)$ is fitted and all the other corrections for wrong tags 
scale with $D^{c}_{\rm PID}(D^0,D_s)$ according to ratios determined with simulated events. 

To summarize, the values for $\dm$, $f_{c}$, $f_{\rm OBC}^{n}$, 
$D^{c}_{\rm PID}(D^0,D_s)$, $f^{n}_{D^0,D_s}$, and 
the widths and relative fractions of the Gaussian components 
for the signal resolution are determined in the likelihood fit. 
The $B$ meson lifetimes are fixed to the values 
quoted in \cite{bib:PDG2000}. 

\begin{figure}[bth] 
\centering 
\includegraphics[width=\linewidth]{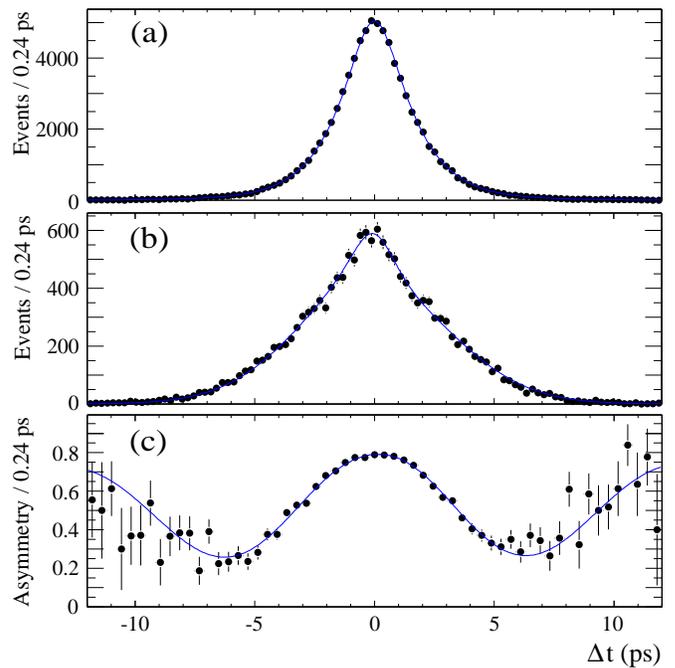} 
\caption{Distributions of decay time difference for (a) opposite-sign and (b) 
same-sign dilepton events; (c) asymmetry between opposite- and same-sign dilepton 
events. Points are data and the lines 
correspond to the fit result.}
\label{alldata}  
\end{figure}

The result of a binned maximum likelihood fit to the data sample with 
the requirement $|\delt|<12$\ps\ yields 
$\dm = 0.488 \pm 0.012$\,ps$^{-1}$ and $f_{c}=0.554 \pm 0.014$. 
Figure~\ref{alldata}a and \ref{alldata}b show the $\delt$ distributions 
for opposite- and same-sign dilepton events, respectively, along with 
the result of the fit. Figure~\ref{alldata}c shows the resulting asymmetry 
as a function of $\delt$. The widths  
of the three Gaussians for the signal resolution function are $0.55\pm 0.09$, 
$1.06\pm 0.23$ and $4.8\pm 0.7$\ps, and the corresponding fractions of events 
are 76\%, 22\% and 2\%. 
The probability to obtain a worse fit is 65\%, evaluated with 
an ensemble of
data-sized experiments that are generated with a parameterized
simulation based on the observed total PDF.
The global fit is also performed on a sample
from full Monte Carlo simulation, where the fitted
results for parameters are consistent with generated values.

%
%
The fit result is found to be stable and consistent
under a variety of choices for free parameters, where fixed values
obtained from Monte Carlo simulation are substituted.

%
%
A summary of the systematic uncertainties is given in Table
\ref{sys_all}, where the total is estimated to be
0.0087\,ps$^{-1}$.
The most important contributions are due to 
the $B$ meson lifetimes, the $\delt$ resolution function, and the modeling of OBC events. 
Varying the $B$ meson lifetimes within their known errors~\cite{bib:PDG2000}
contributes an uncertainty of 0.0064\,ps$^{-1}$ on $\dm$. 

\begin{table}[htbp] 
\centering
  \caption{ Summary of systematic uncertainties}
\small
\begin{tabular}{|l|c|} \hline
Source   & $\sigma(\dm)$ 
ps$^{-1}$ \\ \hline\hline
   $B$ lifetimes                    & $0.0064$ \\
   OBC resolution/lifetimes         & $0.0026$ \\
   $\delt$ dependence of resolution & $0.0043$ \\
   $z$ scale and SVT alignment      & $0.0020$ \\
   OBC fractions/wrong tags         & $0.0020$ \\ 
   Hadron misidentification         & $0.0010$ \\ 
   $J/\psi$ fraction                & $0.0003$ \\
   Continuum parameterization       & $0.0009$ \\
   Binned fit bias                  & $0.0006$ \\ 
   Beam energy uncertainty          & $0.0005$ \\ \hline
   Total                            & $0.0087$ \\ \hline
  \end{tabular}
  \label{sys_all}
\end{table}
The systematic error due to the uncertain knowledge of the resolution function 
for OBC events is estimated by varying the parameters within
their errors from the fit to simulated events, including the effect
of correlations. A possible scale uncertainty  
between data and simulation is estimated by allowing a conservative increase 
of 20\% in the OBC resolution width. 
The overall uncertainty due to the OBC resolution function is
0.0026\,ps$^{-1}$. 
The impact of our treatment for the $\delt$ dependence
of both the signal and OBC resolution functions and
the boost approximation has been studied with 
large parameterized Monte Carlo samples, which are
based on the observed dependence in full simulation. 
Neglecting the $\delt$ dependence  
results in a bias for $\dm$ of $-0.0045$\,ps$^{-1}$. 
The fit result has been corrected to account for this bias and a corresponding
systematic error of 0.0043\,ps$^{-1}$ is assigned. 
Knowledge of the absolute $z$ scale of the detector and 
the residual uncertainties in the SVT local alignment 
give a combined error of 0.0020\,ps$^{-1}$. 

Systematic effects due to the limited knowledge of the parameters 
of the OBC PDF, which are taken from simulated events, 
are greatly reduced by fitting the fractions of OBC and 
the short-lived charm component in neutral $B$ events. 
The remaining systematic uncertainty (0.0020\,ps$^{-1}$) is estimated 
by varying the otherwise fixed charm-related parameters (the amount of $D_s$, $f^{c}_{D^0,D_s}$ 
and the various fractions of cascades) by 10\%, 
both coherently and in independent directions. This is
a conservative range, given our present knowledge of the physics processes involved. 

The ratios between the various wrong-tag factors  
due to hadron misidentification (PID) are conservatively 
varied by 30$\%$ in the fit. The
maximum effect is obtained when the signal and cascade PID 
wrong-tag corrections are varied in  opposite directions. 
In this case, the total systematic error is 0.0010\,ps$^{-1}$. 

The uncertainty on the fraction of $J/\psi$ is 30\%, which contributes an error
on $\dm$ of 0.0003\,ps$^{-1}$. The effective lifetime, the fraction of 
same-sign events, and the fraction of continuum events are varied independently, 
giving a combined systematic error of 0.0009\,ps$^{-1}$. 
The dependence of the fit result on the number of bins has been estimated 
with a parameterized Monte Carlo simulation. 
A shift of $-0.0006$\,ps$^{-1}$ in $\dm$ 
is observed and a corresponding correction applied
with a systematic error of 0.0006\,ps$^{-1}$.
The uncertainty (0.1\%) on the absolute scale of the  beam 
energies gives an error of 0.0005\,ps$^{-1}$ on $\dm$. 

%
%
%

In conclusion, the neutral $B$ meson oscillation frequency has been measured with an 
inclusive dilepton sample to be 
\begin{equation*}
\dm = 0.493 \pm 0.012 (stat) \pm 0.009 (syst)\,
{\mathrm{ps}}^{-1}. 
\end{equation*}
This result is the single most precise measurement to date
and is consistent with a recent \babar\ measurement with
a fully reconstructed \Bz\ sample~\cite{babar0102}. 

%
%
\input pubboard/acknow_PRL.tex
%
%
\end{document}

%% file: symbols.tex
\def\Dp      {\ensuremath{D^+}}
\def\Dm      {\ensuremath{D^-}}
\def\Dstar   {\ensuremath{D^{*}}}
\def\Dstarp  {\ensuremath{D^{*+}}}
\def\Dstarm  {\ensuremath{D^{*-}}}



%% file: pubboard/authors_0122.tex
%
\author{B.~Aubert}
\author{D.~Boutigny}
\author{J.-M.~Gaillard}
\author{A.~Hicheur}
\author{Y.~Karyotakis}
\author{J.~P.~Lees}
\author{P.~Robbe}
\author{V.~Tisserand}
\affiliation{Laboratoire de Physique des Particules, F-74941 Annecy-le-Vieux, France }
\author{A.~Palano}
\author{A.~Pompili}
\affiliation{Universit\`a di Bari, Dipartimento di Fisica and INFN, I-70126 Bari, Italy }
% ...
\collaboration{The \babar\ Collaboration}
\noaffiliation

%% file: pubboard/acknow_PRL.tex
We are grateful for the excellent luminosity and machine conditions
provided by our \pep2\ colleagues.
The collaborating institutions wish to thank 
SLAC for its support and kind hospitality. 
This work is supported by
DOE
and NSF (USA),
NSERC (Canada),
IHEP (China),
CEA and
CNRS-IN2P3
(France),
BMBF
(Germany),
INFN (Italy),
NFR (Norway),
MIST (Russia), and
PPARC (United Kingdom). 
Individuals have received support from the Swiss NSF, 
A.~P.~Sloan Foundation, 
Research Corporation,
and Alexander von Humboldt Foundation.